# Inactivation cross section induced by heavy ions of different energies in Bacterial cells of *E. coli*: an analytical approach

E. M. Awad [a, *] and M. Abu-Shady [b]

[a,] *Physics Department, Faculty of Science, Menoufia University, Shebin El-Koom, Menoufia 32511 Egypt*

[b,] *Mathematics and Computer Science Department, Faculty of Science, Menoufia University, Shebin El-Koom, Menoufia 32511 Egypt*

[*,]*corresponding author e-mail: ayawad@yahoo.com, awadem704@gmail.com, elsayed_awad@science.menofia.edu.eg (E.M. Awad)*

**Abstract**

An analytical inactivation cross section formula based on the analytical formula for calculating the radial dose distribution of Awad et al. (Awad et al., 2018) was proposed. The formula is a multi-hit model based on the track structure theory of Katz that was not updated since Katz and co-workers in 1970s-1980s. The formula was solved numerically to calculation the inactivation cross section, σ and action cross section for double strand break, $\sigma_{DSB}$ for the two strains of the vegetative *E. coli* cells ($B_{S-1}$ and B/r) bombarded with heavy ions of O, Ne, Ar, Kr, Xe, Au, Pb and U of energy between 1.1 to 19.4 MeV/n. The response of the cell and hence the corresponding inactivation cross section is affected by m, the number of targets that must be inactivated per cell; $D_0$, the characteristic x or γ-ray dose; $a_0$, the target radius and $R_{max}$, the maximum range travelled by the liberated δ-ray in the medium. The variations in the inactivation cross section in terms of these parameters were studied. The model predicts the available measured inactivation cross sections data for *E. coli* cells ($B_{S-1}$ and B/r), σ as well as the double strand break, $\sigma_{DSB}$ ones with good accuracy. It was found that as the ion charge increases as the inactivation probability increases and the corresponding inactivation cross section of the ion increases.

**Introduction**

Effect of radiation on DNA and biological tissues is old/new subject and is always gain interest in the past, present and future. In radiation biology, the amorphous track model was introduced in 1960s of the last century (Butts and Katz,1967; Kobetich and Katz, 1968; Katz et al., 1971) and its reduction model, the local effect model (LEM) by Scholz and Kraft (Scholz and Kraft, 1994; Scholz and Kraft, 1996; Scholz et al., 1997) were used to calculate the biological effect of charged particles. The two biophysical models are widely accepted and have been used to calculate the effect of radiation on cells and inactivation cross section of ions to enzymes and viruses in the last few decades. The two approaches are almost similar and the difference between them mostly lies in the concept of local dose (Scholz and Kraft) and average dose (Katz) and the definition of the target size. In LEM cell inactivation is assumed to be due to production of lethal events of certain number in a specific sensitive location (the cell nucleus). No interactions of sub lethal damages over large distances of the order of micrometers are required to produce lethal events. The present calculations, however, is based on the ion kill-gamma kill model of Katz. In which, the effects produced by secondary electrons from γ-rays and those from secondary electrons from heavy ion (δ-rays) are comparable at the same dose. Two radiation actions were incorporated, γ-kill and ion-kill and this makes it is possible to explain the difference between the effect of low and high-LET radiation.

Modeling and predicting the radial dose D(r,R) as a function of the radial distance, r from the ion pass and δ-ray range, R in a given medium is essential for calculating the ion inactivation cross section. The radial dose distribution due to ion interaction with a given medium, D(r,R) was first introduced by Butts and Katz in the 1960s (Butts and Katz, 1967; Kobetich and Katz, 1968) using a simple empirical electron range-energy relationship. This pioneer approach by Katz and co-workers led to formulate many biophysical models. It is worth noting that calculating D(r,R) models had many modifications and improvement during the last few decades (Katz, 1978; 1983; 1985; Zhang et al., 1985; Waligórski et al., 1986; Kiefer and Straaten, 1986; Cucinotta et al., 1995; Zhang et al., 1994; Chan and Kellerer, 1997; Cucinotta et al., 1997; Cucinotta et al., 1999, Korcyl, 2012; Waligórski et al., 2015 (a,b) and Awad et al., 2018). For more information, please refer to these references and references therein. Radial dose, D(r,R) is utilized for radiation transport software (Bernal et al., 2015), developing a treatment planning system and estimating the cell survival rate in the treatment planning system for heavy particle cancer therapy (Waligórski et al., 2015-b; Nikjoo et al., 2016) and predict the cell surviving rate and the relative biological effectiveness (RBE) of the ions (Waligórski et al., 2015-a).

Measuring the inactivation cross section, σ induced by heavy ions on viruses, bacteria and DNA were carried out by many authors in the literature. For SV-40 virus in EO buffer, σ was carried out experimentally as well as theoretically (Katz and Wesely; 1991). Inactivation and mutagenic effect in spores of *Bacillus subtilis* (*B. Subtilis*) irradiated with heavy ions were investigated by Baltschukat and Horneck (Baltschukat and Horneck, 1991). Dose dependent decline in surviving fraction in Chines hamster V79 cells exposed to high LET ions has been studied by Pathak et al. (Pathak et al., 2007). Experimental and theoretical cross sections for *E. coli* mutants B,

B/r, and $B_{S-1}$ after irradiated by Heavy-Ion Irradiation were studied (Katz and Zachariah, 1993). Inactivation cross section of ions for dry enzymes and viruses were calculated using an empirical formula (Liu et al., 1994 and Liu et al., 1996). Heavy ions induced inactivation and mutagenic effect in microorganisms such as yeast cells which provide astronauts needs in active space exploration was studied (Wang el al., 2010). Quantitative analysis of radio-induced DNA damage exposed to protons at Bragg-peak energy was given by Souici et al., (Souici et al., 2017).

The aim of the present work is to calculate the inactivation cross section, σ using a new analytical formula. This formula is based on the analytical approach proposed by Awad et al. (Awad et al., 2018) for calculating the radial dose distribution D(r,R). The formula is a multi-hit model based on the track structure theory of Katz and is considered as an alternative to Monte Carlo code for calculating the radial dose distribution in a given medium due to the passing charged ion. The calculations were carried out for two strains of Bacterial cells of *E.Coli.* Estimating the inactivation cross sections, σ for the two strains ($B_{s-1}$ and B/r) *E. coli* cells bombarded with heavy ions of O, Ne, Ar, Kr, Xe, Au, Pb and U of energy between 1.1 to 19.4 MeV/n were carried out. In addition to calculate the induction double strand break cross section, $\sigma_{DSB}$ for those two strains. The different parameters that affect σ of m, $D_0$, $a_0$ and $R_{max}$ were studied. The calculated inactivation cross sections were as compared to the measured ones. The inactivation probability for each ion was studied.

**Experimental data**

The present calculations were compared to data that obtained from a well-designed experiment in which different ions of different charges with almost similar energy/n were utilized to irradiate the samples. Samples of two strains ($B_{S-1}$ and B/r) of the vegetative cells of *Escherichia coli (E. coli)* of almost similar DNA structure were used (Schäfer et al., 1994). They are considered as a simple test system to measure the induction of DNA double strand breaks (DSB) and their correlation with cell survival dependent on the radiation quality. The two *E. coli* strains were bombarded with heavy ions of Ne, Ar, Kr, Au, Pb and U. The inactivation cross section and double strand break cross section were measured. Another inactivation cross section data for $B_{S-1}$ and B/r *E. coli* bombarded with O, Ne and Xe were obtained (Schäfer et al., 1987) were added. Table 2 compile the different ions of energy between 1.1 to 19.4 MeV/n quoted in the present work and there references. Schäfer et al 1994 data will be superimposed in the present work as (Exp), while Schäfer et al., 1987 data will be quoted as (Exp report).

**Methodology**

Inactivation (loss of colony forming ability) of enzymes, bacteria and viruses by energetic heavy ions gained large interest and still, therefore, many approaches were suggested for calculating the inactivation cross section. Calculating inactivation probability, P and σ is mainly depends on the $\frac{D(r,R)}{D_0}$ ratio. For each D(r,R) formula there is inactivation cross section formula. Therefore, there are Katz (Butts and Katz; 1867), Zhang (Zhang et al., 1985); Waligórski (Waligórski et al., 1987); Cucinotta (Cucinotta et al., 1997) as well as Waligórski (Korcyl, 2012; Waligórski et al., 2015) approaches. The present calculations (will call it Calc) work will be compared with Katz and Zhang calculations (Zhang et al., 1985) and will call it (Katz and Zhang).

The multi-hit inactivation cross section is a statistical model introduced by Katz (Katz et al., 1971; Katz et al., 1996). In that model it was assumed that the effects produced by secondary electrons from γ-rays and those from secondary δ-electrons from heavy ions are comparable at the same dose. Therefore, the probability for producing effects in a macroscopic volume by a given dose of γ-rays in the target was used to estimate the probability of producing effects in that target by a charged ion as follows:

$$P = 1 - e^{-\left(\frac{D(r,R)}{D_0}\right)} \qquad (1)$$

where, $D_0$ is the γ-ray dose for 37% survival of the cell and $D(r,R)$ is the average radial dose over the target volume. $D(r,R)$ is assumed to be the energy deposited per unit mass in a short cylinder of radius $a_0$, whose axis is parallel to the path of the ion. A semi-empirical analytical model based on electronic radiation damage is introduced by (Awad et al., 2018) for calculating the radial dose distribution $D(r,R)$, In this approach, the empirical electron range-energy formula of Tabata et al. (Tabata et al., 1972) was considered where the electron range, *R* in cm is given as:

$$R(T) = a_1 \left( \frac{1}{a_2} \ln\left(1 + a_2 \frac{T}{mc^2}\right) - \frac{a_3 \frac{T}{mc^2}}{1 + a_4 \left(\frac{T}{mc^2}\right)^{a_5}} \right), \qquad (2)$$

*where*

$$a_1 = \frac{b_1 A_T}{\rho Z_T^{b_2}}, \qquad a_2 = b_3 Z_T, \qquad a_3 = b_4 - b_5 Z_T,$$

$$a_4 = b_6 - b_7 Z_T, \qquad a_5 = \frac{b_8}{Z_T^{b_9}},$$

where $Z_T$ and $A_T$ are the atomic and mass number of the target, respectively. $b_i$ (i=1,2,…,9) are constants independent of absorber material and are given in Table 1. $T$ is the energy of the ejected δ-electrons and its maximum value as a function of $\beta$ (the ion velocity in medium relative to velocity of light in vacuum) is given by:

$$T_{max} = \frac{2mc^2 \beta^2}{1-\beta^2}. \qquad (3)$$

The second important step is to calculate the electron energy in MeV as a function of the electron range in cm. The inverse of the energy-range of Eq. (2) was found by Tabata et al. (Tabata, et al., 1972) as follows:

$$T(R) = 0.511 c_1 \left( e^{\frac{R}{c_1}(c_2 + \frac{c_3}{1+c_4 R^{c_5}})} - 1 \right), \qquad (4)$$

where $c_i$ (i=1,2…5) are constant for a given target ($Z_T$, $A_T$) and $d_i$ (i=1,2,…,9) are another constants independent of absorber material and are given in Table 1 as:

$$c_1 = \frac{d_1}{Z_T}, \qquad c_2 = \frac{d_2 Z_T^{d_3}}{A_T}, \qquad c_3 = d_4 - d_5 Z_T,$$

$$c_4 = \frac{d_6}{Z_T^{d_7}}, \qquad c_5 = \frac{d_8}{Z_T^{d_9}}$$

These two steps are important for defining the necessary constants and for more details, please refer to Awad et al., 2018. Finally, the radial dose, $D(r,R)$ depends on r and R and its formula was defined as follows:

$$D(r,R) = \frac{-CZ^{*2}}{\rho 2\pi r c_1 \beta^2} \frac{e^{X_4} X_3 (X_2 - X_1) e^{\left(\frac{R-r}{c_1} X_1\right)}}{(e^{X_4} - 1)^2} \tag{5}$$

Where

$$X_1 = c_2 + \frac{c_3}{1 + c_4(R-r)^{c_5}} \tag{6}$$

$$X_2 = \frac{c_3 c_4 (R-r)^{c_5}}{(1 + c_4(R-r)^{c_5})^2} \tag{7}$$

$$X_3 = \frac{c_2}{c_1} + \frac{\frac{c_3}{c_4}(1 + c_4 R^{c_5}) - \frac{c_3 c_4 c_5}{c_1} R^{c_5}}{(1 + c_4 R^{c_5})^2} \tag{8}$$

$$X_4 = \frac{R}{c_1}\left(c_2 + \frac{c_3}{1 + c_4 R^{c_5}}\right) \tag{9}$$

For water medium, $\rho = 1 \frac{gm}{cm^3}$ and $C$ is a constant (Zhang et al., 1985) equals

$$C = \frac{2\pi N e^4}{mc^2} = 8.5 \frac{keV}{mm} = 1.369 \times 10^{-7} \frac{erg}{cm} = 1.369 \times 10^{-14} \frac{J}{cm} \tag{10}$$

The inactivation probability thus can be determined by using Eq. (5) for D(r, R) and hence, the inactivation cross section, σ for a given ion is given as follows:

$$\sigma = 2\pi \int_{R_{min}}^{R_{max}} \int_1^{r=R_{min}} r \left[1 - e^{-\left(\frac{D(r,R)}{D_0}\right)}\right]^m dr\, dR \tag{11}$$

where, m represents the number of targets in the cell which must be hit and inactivated by radiation to produce an observable effect in such cell. For one hit detectors m = 1 i.e. a single hit in a single target suffices to inactivate the detector. For biological cells m is frequently 2 or more (Katz et al., 1996).

An algorithm was constructed to calculate the double integration of σ as given in Eq. (11) by using *Mid-Point numerical Method* of integration for a given target. For calculating the radial dose, the ion parameters of $Z^*$ (the effective charge number of the ion) which was calculated according to Barkas' formula as $Z^* = Z(1 - e^{-125\beta Z^{-\frac{2}{3}}})$ and $\beta$ were incorporated in the calculation as well as the target parameter of $\rho = 1 \frac{gm}{cm^3}$ where the *E. Coli* density was considered as a water medium. The two integration limits were carried out in such a way that the down limit for the second integration is $R_{min}$ and the upper limit is $R_{max}$ are the limits for radial dose estimation (Eq. 5). For the first integration that is given in Eq. 11, the lower limit is taken as 1 nm and the upper limit is less than or equal $R_{min}$ of the first integration. It is worth noting that, the first integration is the target radius $a_0$.

## 3. Results

### 3.1 Parameters affecting inactivation cross section calculations

The response of the *E. coli* cells ($B_{S-1}$ and B/r) to the bombarded ions was studied through investigating the inactivation cross section of these cells using Eq. 11. Equation 11 was solved numerically to calculation the inactivation cross section, $\sigma$ and induction double strand break cross section, $\sigma_{dsb}$ for the two strains of the vegetative cell of *E. coli* cells ($B_{s-1}$ and B/r) bombarded with heavy ions of O, Ne, Ar, Kr, Xe, Au, Pb and U of energy between 1.1 to 19.4 MeV/n. The data for each strain will be demonstrated and discussed separately later on.

Inactivation cross section depends on four parameters (m, $D_0$, $a_0$ and $R_{max}$). The number of targets that must be hit by radiation to cause an observable effect (inactivate) per cell is m. $D_0$ is the characteristic dose and it is the γ-ray or x-ray dose for 37% of cell survival, $a_0$ is the target radius and finally, $R_{max}$ is the maximum δ-electrons range corresponding to the maximum δ-ray energy transferred due to the energetic charged ion (Eq. 3) and it is the upper limit for the second integration in Eq. 11. In this part, the effect of these parameters on σ will be discussed and inactivating of B/r *E.coli* cells by Argon ion will be presented as an example. The variation in σ with the ion energy at different m values of m=1,2,3,4 and 5 and at $D_0$=61 Gy and target radius, $a_0$=0.2 µm is shown in Fig. 1. One can observe that, the inactive cross section decreases with m and m=2 gives the best fit in this case. It was found that σ is very sensitive to m and m=2 give the best fit for σ for the two E. coli cells (B and B/r) and m=3 give the best fit for the double strand break action cross section, $\sigma_{DSB}$ for E. coli cell DNA.

Inactivation cross section as a function the Ar energy for different target radius of $a_0$=0.1, 0.2, 0.3 and 0.5 µm and at m=2 and $D_0$=61 Gy is illustrated in Fig.2. It can be seen that, at low energy (less than ≈8 Mev/n) the target radius is slightly effective while for energy greater than 8 MeV/n is $a_0$ is not changing σ significantly. It was found that target radius that gives best fit to the experimental data is at $a_0$= 0.2 µm (200 nm). Therefore, all targets radii will be fixed at 0.2 µm.

Regarding the characteristic dose, $D_0$ the inactivation cross section for B/r E. coli cell was studied at different Ar energy and at different $D_0$ values, see Fig. 3. One can observe that $D_0$ values are significantly changed the obtained inactivation cross section and σ is very sensitive to $D_0$. In opposite to previous study, the present work was fitting the inactivation cross section using different $D_0$ values for each ion. The previous studies, were fixing $D_0$ value for all ions. Table 3 is gathering the characteristic dose, $D_0$ for the different ions in comparison with the literature $D_0$ values for both vegetable E. coli cells.

Table 4 as well as Fig.4 demonstrates the effect of $R_{max}$, the upper integration limit on inactivating B/r E. coli cells at different Ar energy. Using the full δ-ray range ($R_{max}$) that was deduced using Eq. 3 and Eq. 2 gives progressive cross section values. It was found that at energy between 2 to 10 MeV/n the deduced $R_{max}$ using Eq. 3 and Eq. 2 fits most of the studied data, however, a problem exist in the region above 10 MeV/n where the calculated σ cannot fit the experimental data at such high energy regions. Therefore, many efforts and by using trial and error method, the present work suggested $R_{max}$ values that fit most of the studied experimental data and is given in Table 4.

*3.2 $B_{S-1}$ E. coli inactivation cross section*

In the energy range 2-20 MeV/n, the present model calculated the inactivation cross section for *$B_{S-1}$ E. coli* cells bombarded with O and Ne ions using the fitting parameters m=2, $a_0$=0.2 µm and $D_0$= 12.5 Gy (see Table 3). σ ($µm^2$) as a function of O and Ne energy was plotted in Fig. 5. Good agreement between the present calculations and the experimental data for O (O, Exp.) and Ne (Ne, Exp.) data and it is closer to Zhang and Katz calculations for O as well. Similarly, the experimental data for *$B_{S-1}$ E. coli* cells strike by Ar and Kr at different energies are well predicted by the present model, Fig.6. Table 3 compiles the implemented model parameters. The present calculations as well as the experimental data are in agreement with calculated data by Katz and Zhang for both ions.

Fig. 7 and Fig. 8 present a comparison between experimental inactivation cross section for *$B_{S-1}$ E. coli* cells bombarded by Pb, Au, Xe and U ions. The fitting parameters for these ions are compiled in Table 3. One can observe that as the ion charge increases as the cross section of inactivation increases. Reasonable matching between the present calculations and the experimental data at both low and high energy part in agreement with Katz and Zhang calculations as well.

*3.3 B/r E. coli inactivation cross section*

The present model is able to predict the inactivation cross section for *B/r E. coli* cells bombarded by heavy ions quit well in comparison with both experimental and Katz and Zhang calculations. It must emphasis that inactivation cross section for *B/r E. coli* cells was carried out using target radius, $a_0$=0.2 µm and m=2. Comparison of the predicted inactivation cross section for *B/r E. coli* cells produced by the present approach and the experimental inactivation of *B/r E. coli* cells bombarded by O and Ne is illustrated in Fig. 9. The corresponding fitting parameters for this strain are given in Table 3. Inspecting Table 3 shows that B/r E. coli cells are more resistant to radiation than the $B_{S-1}$ *E. coli* cells. It was found that $D_0$ for *B/r E. coli* is always larger than $D_0$ for $B_{S-1}$ *E. coli* cells at a given ion and at a given energy.

Inactivation of *B/r E. coli* cells by Argon, Krypton, Xenon, Lead and gold ions as a function of energy/n are shown in Fig. 10 and Fig. 11. General satisfactory agreement has been found among the present calculations, experimental data and Katz and Zhang calculations especially at the low energy part. However, some deviations appeared between calculated and experimental data at high energy part.

*3.4 Action cross section for DSB*

Interaction of radiation with DNA causing double strand break is the major lesions in the genomes, therefore, the action cross section for DNA samples, $\sigma_{DSB}$ for *E. coli* cells ($B_{S-1}$ and B/r) was also calculated using Eq. 11. The beauty of the present approach is its ability to predict the action cross section for double strand break, $\sigma_{DSB}$ for *E. coli* cells ($B_{S-1}$ and B/r) DNA using the same fitting parameters (Table 3 and 5) for a given strain or cell. The only difference is that DNA needs three hits (m=3) while the cell itself is inactivated by two hits only (m=2). DNA samples were bombarded with different ions at different energies. The different fitting parameters for each ion are compiled in Table 5. It was found that the experimental data was well predicted using the target radius, $a_0$=0.2 µm and m=3 and $D_0$ values are different for different ions. In this table, the experimental action cross section for double strand break, $\sigma_{DSB}$ for *E. coli* cells ($B_{S-1}$ and B/r) DNA are compared to the calculated ones. Global agreement between experimental and calculated data was found especially at the low energy ions but at higher energy still some disagreement between calculations and experimental data.

*3.5 Ion's inactivation probability*

The inactivation probabilities for $B_{S-1}$ *E. coli*, as an example, bombarded with a series of different ions were studied. Inactivation probability (Eq. 1) of some charged particles of different atomic number (Ne, Ar, Kr, Au, Pb and U) having similar energy per nucleon was calculated as a function of the radial distance, r normal to the ion pass. Inspecting Fig. 12, one can observe that for light ions i.e. Neon can inactivate the bacteria ≈100% if it hit it within ≈ 1 nm from the ion path while Argon ion at ≈ 2 nm, Krypton at ≈ 5 nm. Heavy elements like lead and U can inactivate the bacteria a 100 % if it hit it within ≈ 15 nm and Au is within 20 nm from the ion projectile.

## 4. Discussion

Inactivation cross section of two E. coli strains ($B_{S-1}$ and B/r) by heavy elements was calculated by integration the inactivation cross section, Eq. 11 numerically by the mid-point method. Four parameters are affecting the inactivation cross section values i.e. σ ($a_0$, m, $D_0$, $R_{max}$). It was found that inactivation cross section is slightly depended on the target radius $a_0$ and $a_0$ equal 200 nm or 0.2 μm fits the experimental data for all ions. Therefore, $a_0$=0.2 μm kept constant for σ and $σ_{DSB}$ through-out the present calculations for $B_{S-1}$ and B/r cells. This target radius is comparable to the radius of the cell where E. coli bacterium is about 1–3 μm long and 1 μm width and about 0.25 μm in diameter. The target radius, $a_0$ equal 0.2 μm is closer to previous studies, see Table 3.

The number of targets that must be hit by radiation to cause an observable effect (inactivate) per cell was found equal 2 (m=2) by the present calculation for E. coli cells ($B_{S-1}$ and B/r). In the meantime, in calculating the action cross section for DNA for the same E. coli ($B_{S-1}$ and B/r) cells, $σ_{DSB}$ it was found that m=3 give the best fit to the experimental data. Therefore, E. coli cells ($B_{S-1}$ and B/r) are not 1-hit detector (m=1) as Katz calculated (Katz, 1996), instead it is a multi-hit detector as expected before by Caucinotta et al., 1996 and Katz 1996 for biological systems. This implies that two electrons passing through the bacterium ($B_{S-1}$ and B/r) and three electrons for their DNA are capable of inactivating E. coli cells. E. coli cells are multi-hit detector is more reasonable compared to the notion that it's one-hit system (a single electron passing through the bacterium is capable of inactivating the cell). For biological cells m is frequently 2 or more (Katz et al., 1993 and Katz et al., 1996). The present finding of m=2 is more reasonable in this regard and suggest the significant of the direct interaction between the incoming ions and cells rather than the indirect interaction.

The characteristic dose, $D_0$ is significantly affecting the calculated inactivation cross section; σ. Katz and co-workers were assuming fixed $a_0$ and $D_0$ values but in the present work $a_0$ was fixed at 0.2 µm while $D_0$ was varying with different ions. Table 3 and compiles the $D_0$ values for E. coli strains and its DNA for the different bombarded ions. One can observe that $D_0$ increases when the atomic number increases as stated in Tables 3 and 5. It was found that the results of the present calculations depend on $D_0$ than the target radius, $a_0$ in agreement with what was observed by Waligórski et al., 1987. Fitting the experimental data, suggested the change in $D_0$ gives better prediction to the experimental data rather than fixing it as previous investigator do. B/r E. coli is more resistant to radiation compared to $B_{S-1}$ strain where $D_0$ for it is always large as illustrated in the tables.

Inspecting the figures as well as the tables, one may observe that the calculated inactivation cross section for heavy ions is not 100% in agreement with the experimental data. One of the probable reasons for this disagreement could be the mean dose distribution over the target radius when particle are passing the target area (Schäfer et al., 1994 and Katz et al., 1996). This assumption is a fundamental one suggested by Katz in his multi-hit model. In comparison with experimental data, the inactivation cross section has the difficulty due to determining the real maximum range, $R_{max}$ travelled by δ-rays (inaccurate determination for $R_{max}$). $R_{max}$ is the upper integration limit and it is significant in both calculating the radial dose (Eq.5) and calculating σ (Eq. 11). Therefore, calculating the maximum δ-ray range by Eq. 3 and Eq. 2 give good fit to the experimental data up to 10 MeV/n. But larger than 10-12 MeV/n, there are some disagreement between the experimental and the calculated data. Therefore, the trial and error method was applied and the suggested $R_{max}$ that improved the model prediction and reasonable agreement with experimental data was achieved. This new $R_{max}$ can be accepted in the view that electrons are not moving in straight line and it suffers from multiple scattering in its path. One important reason for this slightly disagreement at higher energy parts may be the scattering in the experimental data itself. In addition to this, Takahashi et al., 1983 in his attempt to explain the disagreement between experimental and calculated data (Takahashi et al., 1983) suggested that closer to the ion path (10 Å) most atoms or molecules are ionized in such a way that such densely excited or ionized species may react with each other and cause thermal spike effects. Such thermal effect may cause derangement of biological systems and produces boost growing of inactivation cross sections within this narrow region closer to the ion. Therefore, the mean dose distribution over the whole target radius assumption may not exactly hold.

## 5. Conclusion

The present approach is a promising one for calculating the inactivation cross section for bacteria and viruses where good global agreement between the present calculations and the experimental data was obtained.

Inactivation cross section calculation and cell death probability due to radiation is the target of this work and it was found that ions with higher atomic numbers produce larger inactivation probability and inactivation cross section to the E. coli cells ($B_{S-1}$ and B/r) and its DNA compared to the light ions.

E. coli cells are a multi-hit detector where m=2 for the two E. coli strains and m=3 for its DNA rather than a 1-hit detector.

Calculating the inactivation cross section for *B/r E. coli* cells shows that *B/r E. coli* is more resistant or less sensitive to radiation than the $B_{S-1}$ *E. coli* cells where the characteristic dose $D_0$ for *B/r E. coli* is always larger than $D_0$ for $B_{S-1}$ *E. coli* cells at a given ion.

Calculated inactivation cross section is slightly depending on the target radius, $a_0$. In the meantime, it is strongly depend on the chosen $R_{max}$, the travelled δ-ray range in the medium.

The present model predict inactivation cross section for lighter ions (O, Ne, Ar and Ke) very well but for heavier ions (Xe, Au, Pb and U) it fit the experimental data very well at lower energies (less than 10 MeV/n) but it found some difficulties to predict the data at higher energies. The scattering in the experimental data helped to increase this observation. It may be attributed to inaccurate $R_{max}$ determination and the condensed thermal spike region produced by the heavy ions.

In progress, calculating the inactivation cross section for the more complicated mammalian cells; animal and culture cell using the current approach. Estimating the relative biological effectiveness, RBE for the different ions as well as using this method for developing treatment radiotherapy plans for some ions like carbon is under investigations.

Zhang, C. X., Dunn, D.E., Katz, R., 1985. Radial distribution of dose and cross-sections for the inactivation of dry enzymes and viruses. Radiat. Protect. Dos. 13 (1-4), 215-218.

Zhang, C.X., Liu, X.W., Li, M.F., Luo, D.L., 1994. Numerical calculation of the radial distribution dose around the path of a heavy ion. Radiat. Prot. Dosim. 52 (1-4), 93.

Wang, J., Lu, D., Wu, X., Sun, H., Ma, S., Li, R., Li, W., 2010. Inactive and mutagenic effects induced by carbon beams of different LET values in a red yeast strain. Nucl. Instr. and Meth. B, 268, 2719–2723.

**Figure Captions**

Fig.1　　Calculated inactivation cross section at different m values for B/r E. coli cells bombarded by Argon at different energies in comparison with experimental data.

Fig.2　　Calculated inactivation cross section at different target radii, $a_0$ for B/r E. coli cells bombarded by Argon at different energies in comparison with experimental data.

Fig.3　　Calculated inactivation cross section at different characteristic dose, $D_0$ for B/r E. coli cells bombarded by Argon at different energies in comparison with experimental data.

Fig.4　　δ-ray maximum range, $R_{max}$ effect on the calculated inactivation cross section for B/r E. coli cells bombarded by Argon at different energies in comparison with experimental data.

Fig.5　　Inactivation cross section, σ ($\mu m^2$) of $B_{S-1}$ E. coli cells by O and Ne ions at energy 2-20 MeV/n in comparison with experimental data as well as Katz and Zhang calculations.

Fig.6　　Inactivation cross section, σ ($\mu m^2$) of $B_{S-1}$ E. coli cells by Ar and Kr ions at energy 2-20 MeV/n in comparison with experimental data as well as Katz and Zhang calculations.

Fig.7 Inactivation cross section, σ (μm$^2$) of B$_{S-1}$ E. coli cells by Pb and Au ions at energy 2-20 MeV/n in comparison with experimental data as well as Katz and Zhang calculations.

Fig.8 Inactivation cross section, σ (μm$^2$) of B$_{S-1}$ E. coli cells by Xe and U ions at energy 2-20 MeV/n in comparison with experimental data as well as Katz and Zhang calculations.

Fig.9 Inactivation cross section, σ (μm$^2$) of B/r E. coli cells by O and Ne ions at different energies in comparison with experimental data as well as Katz and Zhang calculations.

Fig.10 Inactivation cross section, σ (μm$^2$) of B/r E. coli cells by Ar and Kr ions at different energies in comparison with experimental data as well as Katz and Zhang calculations.

Fig.11 Inactivation cross section, σ (μm$^2$) of B/r E. coli cells by Xe, Pb and Au ions at different energies in comparison with experimental data as well as Katz and Zhang calculations.

Fig.12 Inactivation probability of B$_{S-1}$ E. coli cells bombarded by Ne (10.48 MeV/n), Ar (10.49 MeV/n), Kr (11.22 MeV/n), Au (10.33 MeV/n), Pb (12.75 MeV/n) and U (13.52 MeV/n) ions as a function of the radial distance, r from the ion trajectory.

**Table Captions**

Table 1    Values of the constants $b_i$ and $d_i$ used for electron range-energy equation, Eq. (1) and its inverse relation of energy-range equation, Eq. (2).

Table 2    The experimental data are mainly from Schäfer et al., 1994 (Exp.) and Schäfer et al., 1987 (Exp. report) that will be marked as $^*$.

Table 3    The implemented characteristic dose, $D_0$ for the different ions in comparison with literature value for $B_{S-1}$ and $B/r$ E. coli cells.

Table 4    The upper integration limit, $R_{max}$ where the lower integration limit $R_{min} = 1$ nm.

Table 5    Experimental inactivating Double Strand Break cross section ($\sigma_{DSB}$) data of *E. Coli* cells ($B_{S-1}$ and B/r) bombarded by different ions in comparison the calculated $\sigma_{DSB}$ ones using m=3.

Table 1

| $i$ | $b_i$ | $d_i$ |
|---|---|---|
| 1 | 0.2335 ± 0.0091 | (2.98 ± 0.3)x10$^3$ |
| 2 | 1.209 ± 0.015 | 6.14 ± 0.29 |
| 3 | (1.78 ± 0.36)x10$^{-4}$ | 1.026 ± 0.02 |
| 4 | 0.9891 ± 0.001 | (2.57 ± 0.12)x10$^2$ |
| 5 | (3.01 ± 0.35)x10$^{-4}$ | 0.34 ± 0.19 |
| 6 | 1.468 ± 0.09 | (1.47 ± 0.19)x10$^3$ |
| 7 | (1.18 ± 0.097)x10$^{-2}$ | 0.692 ± 0.039 |
| 8 | 1.232 ± 0.067 | 0.905 ± 0.031 |
| 9 | 0.109 ± 0.017 | 0.1874 ± 0.0086 |

Table 2

| Ion | Z | E (MeV/n) | Ion | Z | E (MeV/n) |
|---|---|---|---|---|---|
| O | 8 | 8* | Kr | 36 | 13.22 |
|  |  | 5* |  |  | 11.22 |
| Ne | 10 | 14.09 |  |  | 6.24 |
|  |  | 11.77 |  |  | 1.1* |
|  |  | 10.50 |  |  | 3* |
|  |  | 4.43 |  |  | 9* |
| Ar | 18 | 19.4* |  |  | 17* |
|  |  | 14.36 | Au | 79 | 5.1 |
|  |  | 13.59 |  |  | 10.33 |
|  |  | 11.4* | Pb | 82 | 15* |
|  |  | 10.50 |  |  | 12.75 |
|  |  | 7.6 |  |  | 11* |
|  |  | 6.7* |  |  | 4.28 |
|  |  | 4.7 | U | 92 | 1.5* |
|  |  | 1.6* |  |  | 4* |
| Xe | 54 | 2.3* |  |  | 9* |
|  |  | 4* |  |  | 13.52 |
|  |  | 11.4* |  |  | 15* |
|  |  | 16.2* |  |  |  |

Table 3

| Literature | | | | Present | | | | |
| --- | --- | --- | --- | --- | --- | --- | --- | --- |
| $D_0$ (Gy) | | | | $D_0$ (Gy) | | | | |
| References | $a_0$(μm) | $B_{S-1}$ | B/r | $a_0$=0.2 μm | $B_{S-1}$ | | B/r | |
| | | | | Z | σ | $σ_{DSB}$ | σ | $σ_{DSB}$ |
| Takahashi et al., 1986 | | 12.6 | 36.5 | 8 | 12.5 | | 20 | |
| Schafer et al., 1994 | 0.2/0.4 | 15.4 | 47.6 | 10 | 20 | | 29 | |
| Zhang and Katz, 1995 | 0.5 | 13.8 | 44.6 | 18 | 42 | | 61 | |
| Katz et al., 1996 | 0.1 | 12.6 | 40 | 36 | 70 | | 80 | |
| Liu et al., 1996 | | 12.6 | 47.6 | 54 | 77 | | 110 | |
| Korcyl, 2012 | 0.1 | 12.6 | | 79 | 110 | | 185 | |
| . | | | | 82 | 120 | | 170 | |
| | | | | 92 | 100 | | | |

Table 4

| E/n (MeV/n) | $R_{max}$ (nm) (Eq. 3 & Eq. 2) | $R_{max}$ (nm) present work |
|---|---|---|
| 2 | 660 | 660 |
| 4 | 1930 | 1930 |
| 6 | 3800 | 3800 |
| 8 | 6240 | 6240 |
| 10 | 9240 | 9240 |
| 12 | 12800 | 11500 |
| 14 | 16800 | 12800 |
| 16 | 21400 | 12800 |
| 18 | 26400 | 12800 |
| 20 | 31600 | 12800 |

Table 5

| Z | Energy (MeV/n) | $D_0$ (Gy) | | DSB | | | |
|---|---|---|---|---|---|---|---|
| | | $B_{S-1}$ | B/r | $B_{S-1}$ | | B/r | |
| | | | | $\sigma_{Exp}$ (µm$^2$) | $\sigma_{Calc}$ (µm$^2$) | $\sigma_{Exp}$ (µm$^2$) | $\sigma_{Calc}$ (µm$^2$) |
| 10 | 14.09 | 20 | 29 | 0.41±0.04 | 0.29 | 0.26±0.08 | 0.13 |
| | 11.77 | 20 | | 0.36±0.06 | 0.42 | | |
| | 10.48 | 20 | 29 | 0.44±0.08 | 0.41 | 0.10±0.07 | 0.17 |
| | 4.43 | 20 | 29 | 0.41±0.14 | 0.35 | 0.31±0.08 | 0.16 |
| 18 | 14.36 | 42 | 61 | 0.66±0.06 | 0.57 | 0.36±0.06 | 0.31 |
| | 13.59 | 42 | 61 | 0.70±0.11 | 0.65 | 0.40±0.11 | 0.29 |
| | 10.49 | 42 | | 0.81±0.12 | 0.73 | | |
| | 7.6 | 42 | 61 | 0.51±0.20 | 0.65 | 0.35±0.27 | 0.32 |
| | 4.7 | 42 | 61 | 0.63±0.13 | 0.52 | 0.73±0.15 | 0.26 |
| 36 | 13.22 | 70 | | 0.72±0.18 | 2.12 | | |
| | 11.22 | 70 | 80 | 1.58±0.16 | 2.35 | 1.78±0.08 | 1.72 |
| | 6.24 | 70 | 80 | 1.19±0.09 | 1.48 | 1.29±0.09 | 1.16 |
| 79 | 10.33 | 110 | 185 | 3.06±0.46 | 9.35 | 3.49±0.33 | 3.94 |
| | 5.1 | 110 | 185 | 2.81±0.67 | 3.85 | 2.54±0.47 | 1.62 |
| 82 | 12.75 | 120 | 170 | 4.10±0.55 | 6.20 | 3.25±0.81 | 3.42 |
| | 4.28 | 120 | 170 | 3.20±0.45 | 1.90 | 2.22±0.43 | 1.10 |
| 92 | 13.52 | 100 | | 4.59±0.97 | 10.5 | | |

Fig. 1

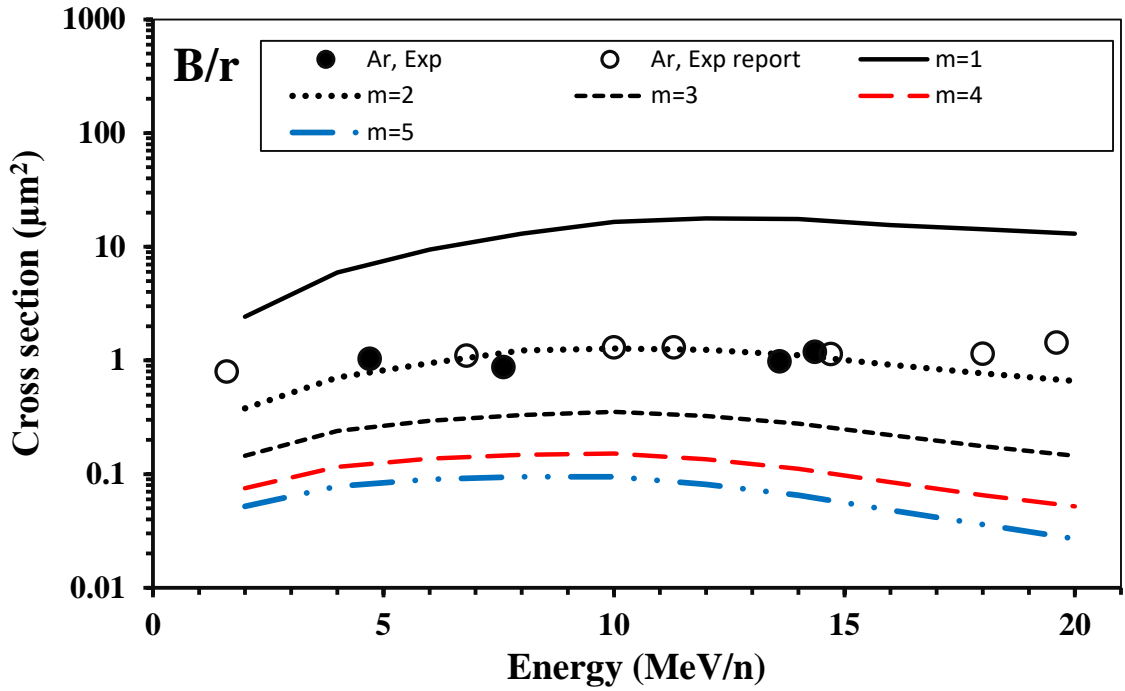

Fig 2

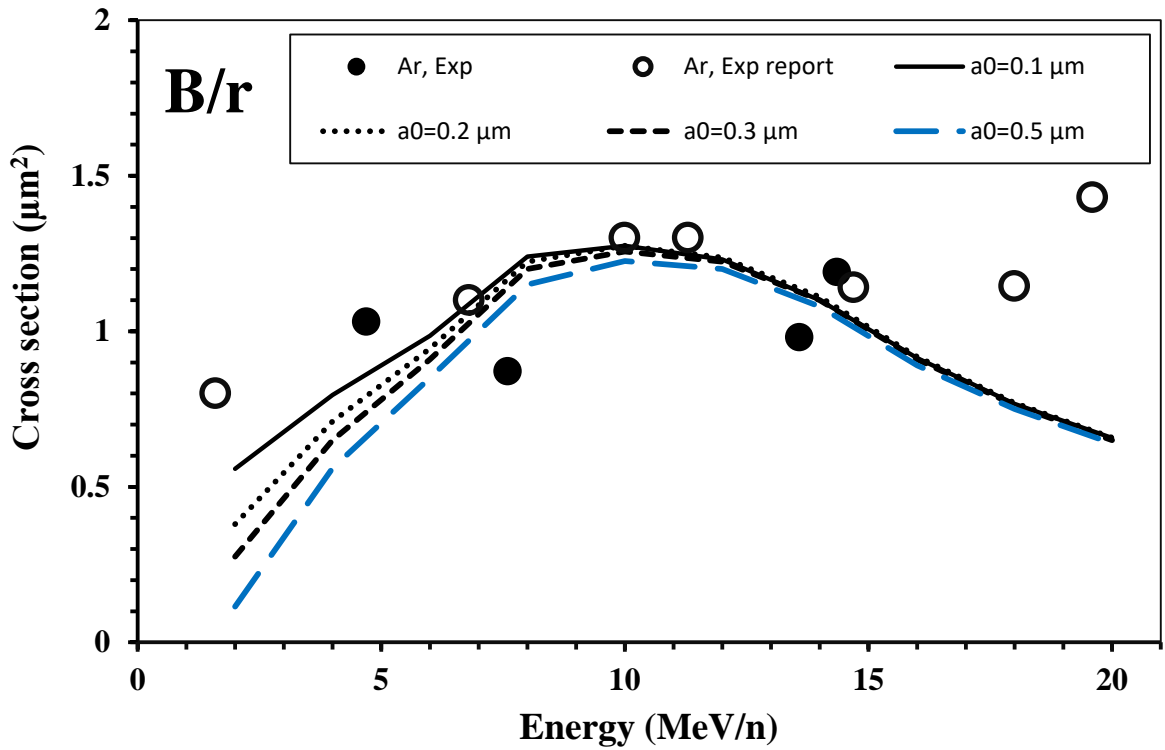

Fig. 3

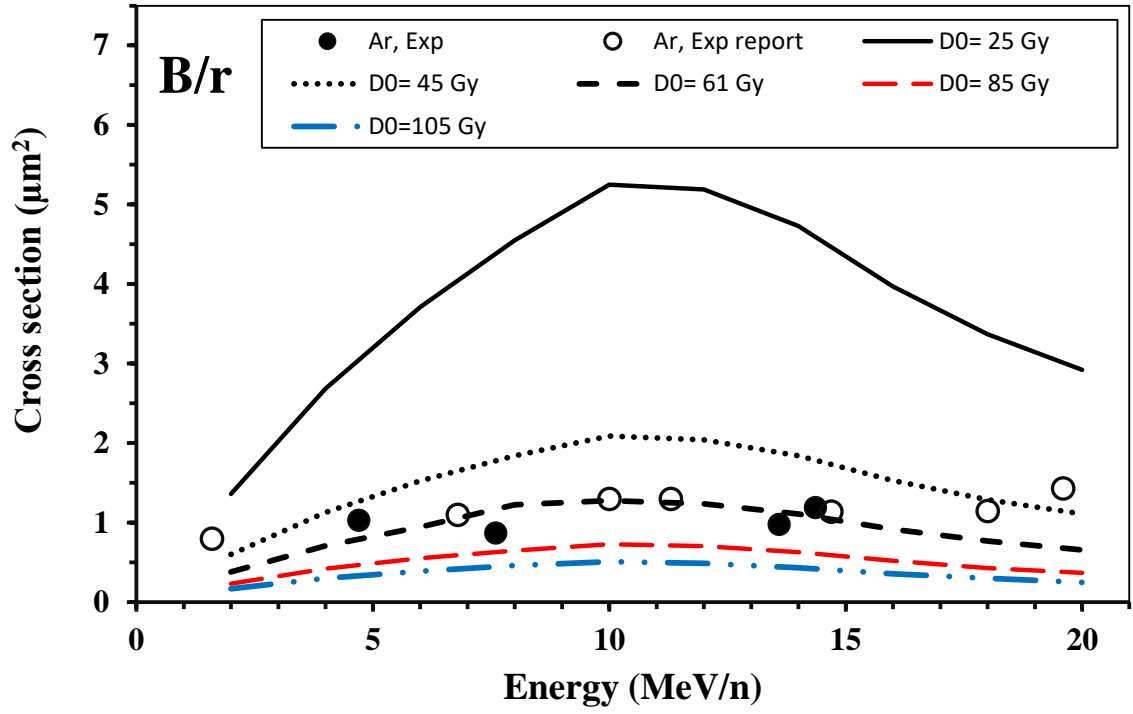

Fig 4

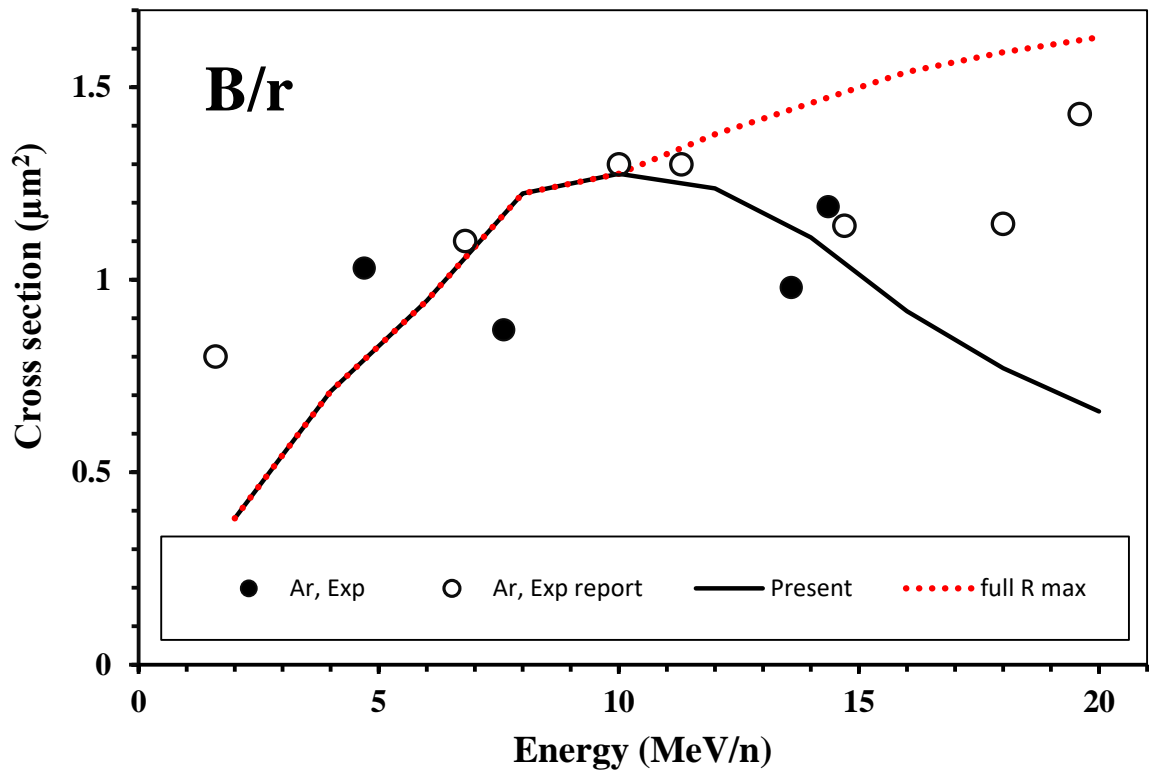

Fig 5

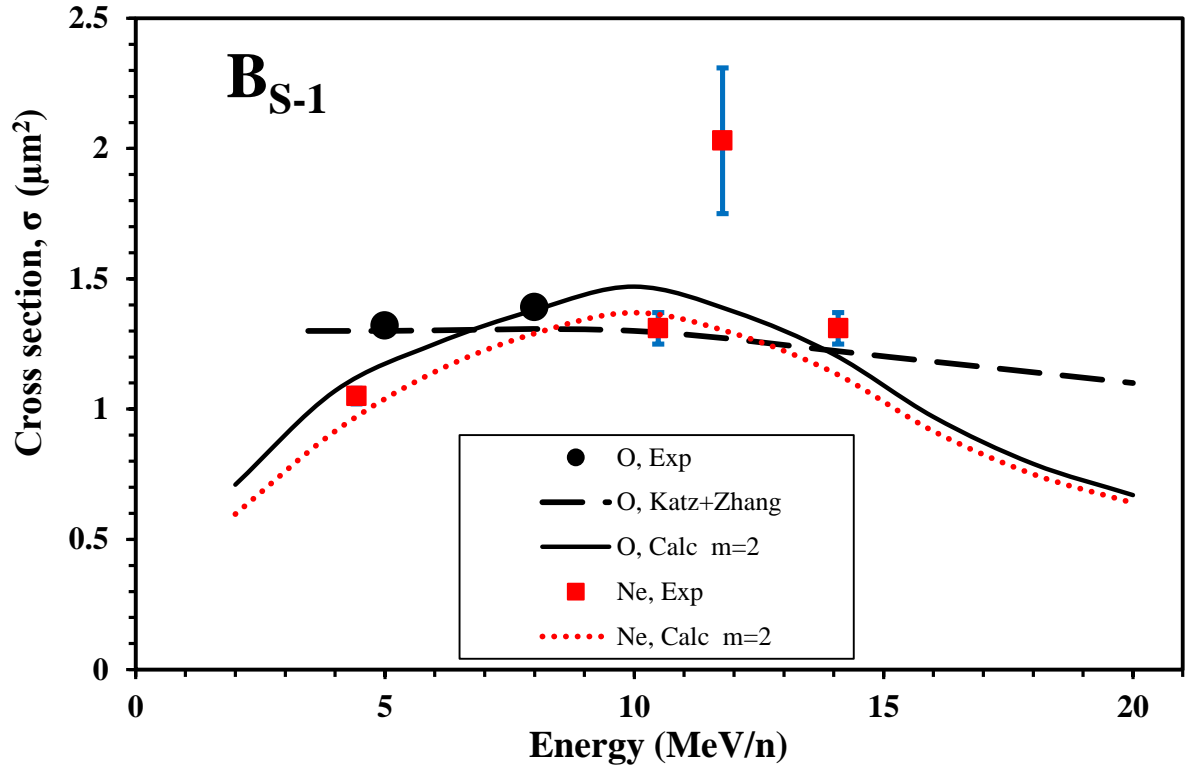

Fig 6

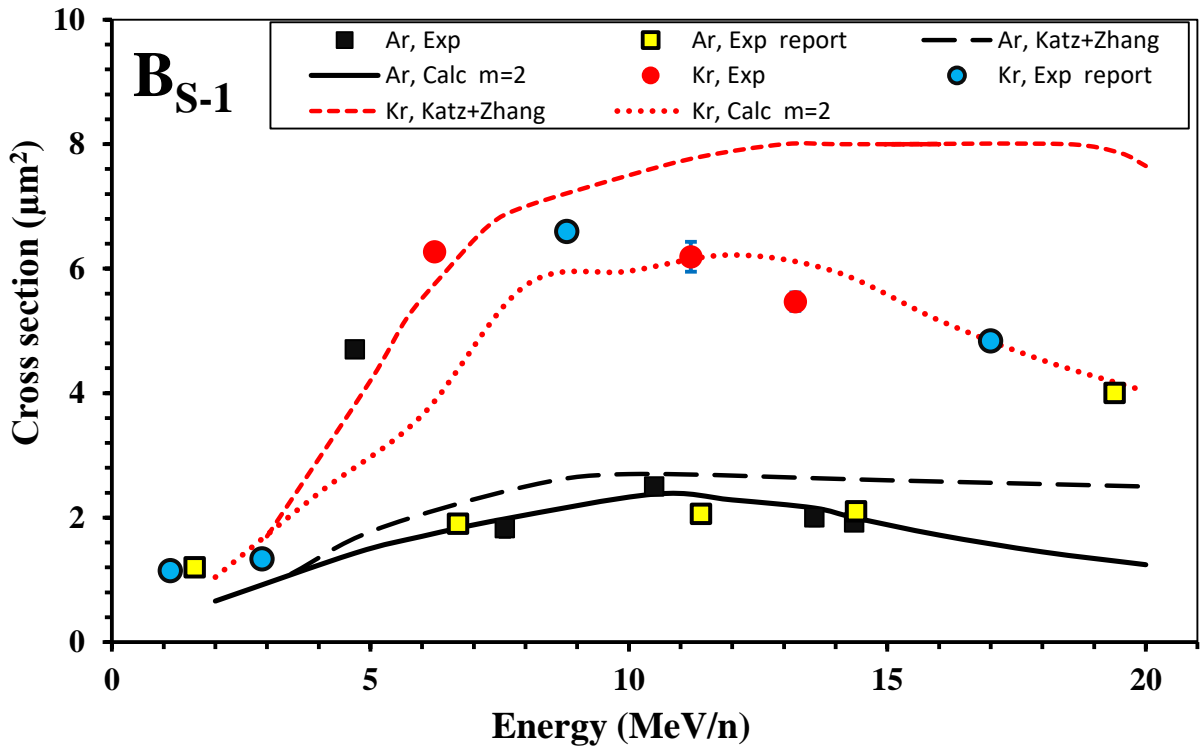

Fig 7

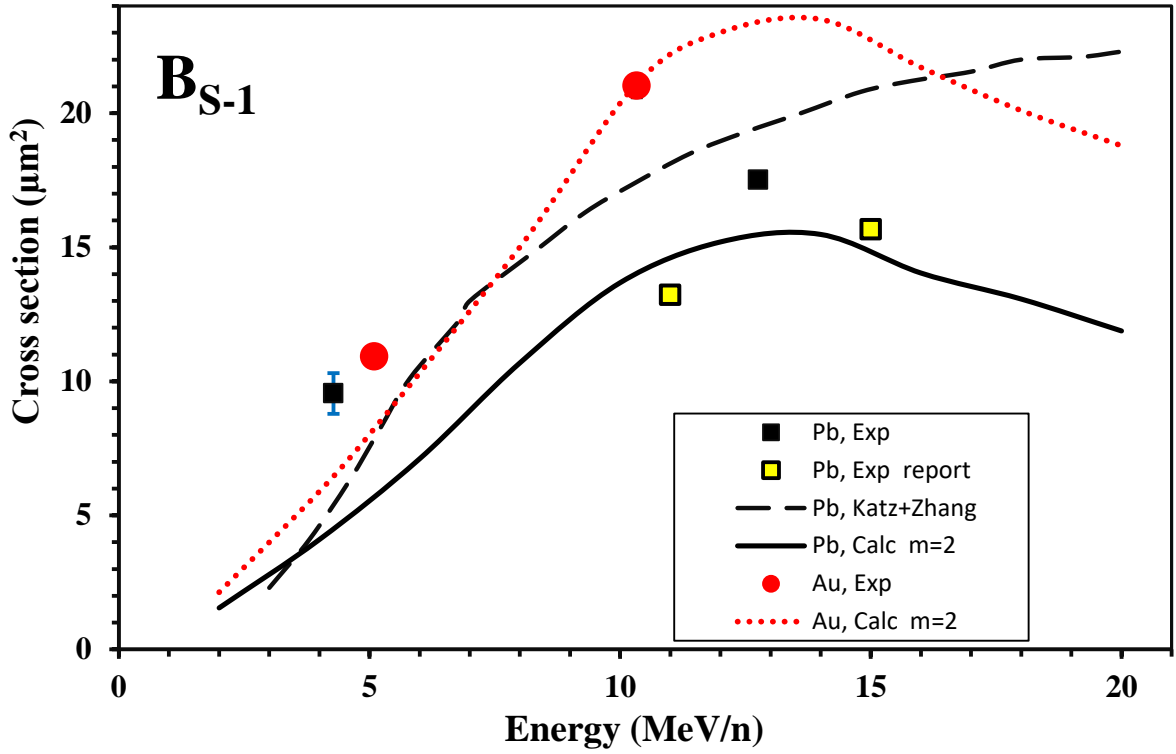

Fig 8

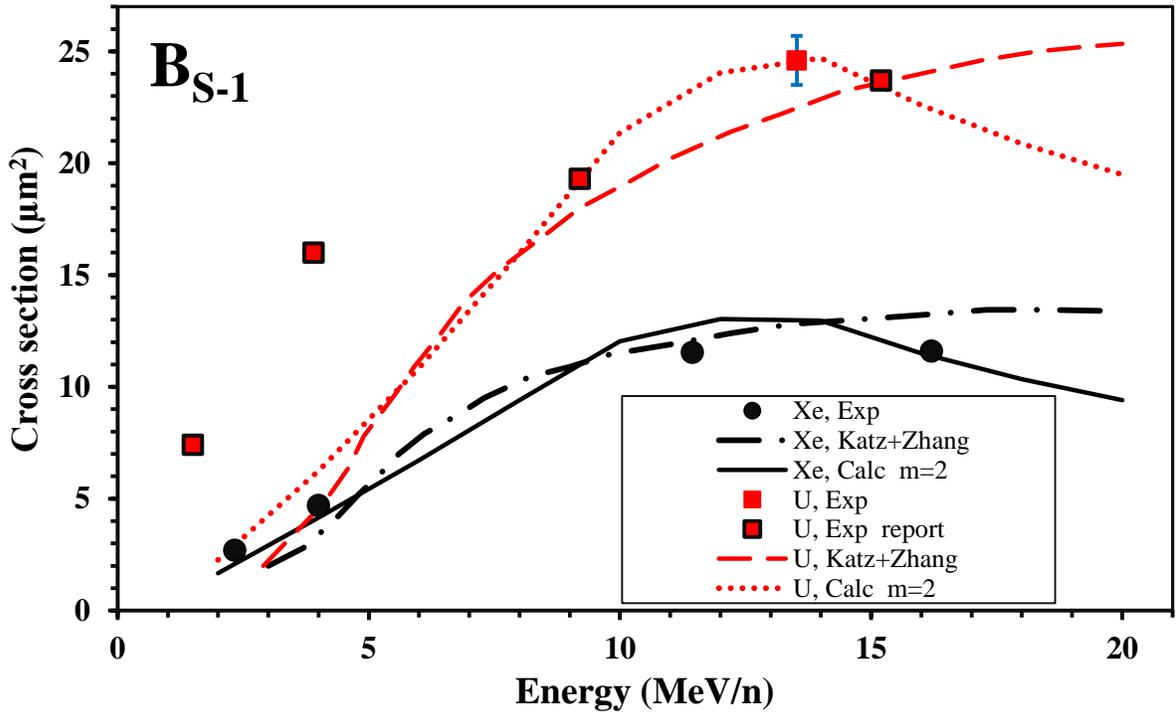

Fig.9

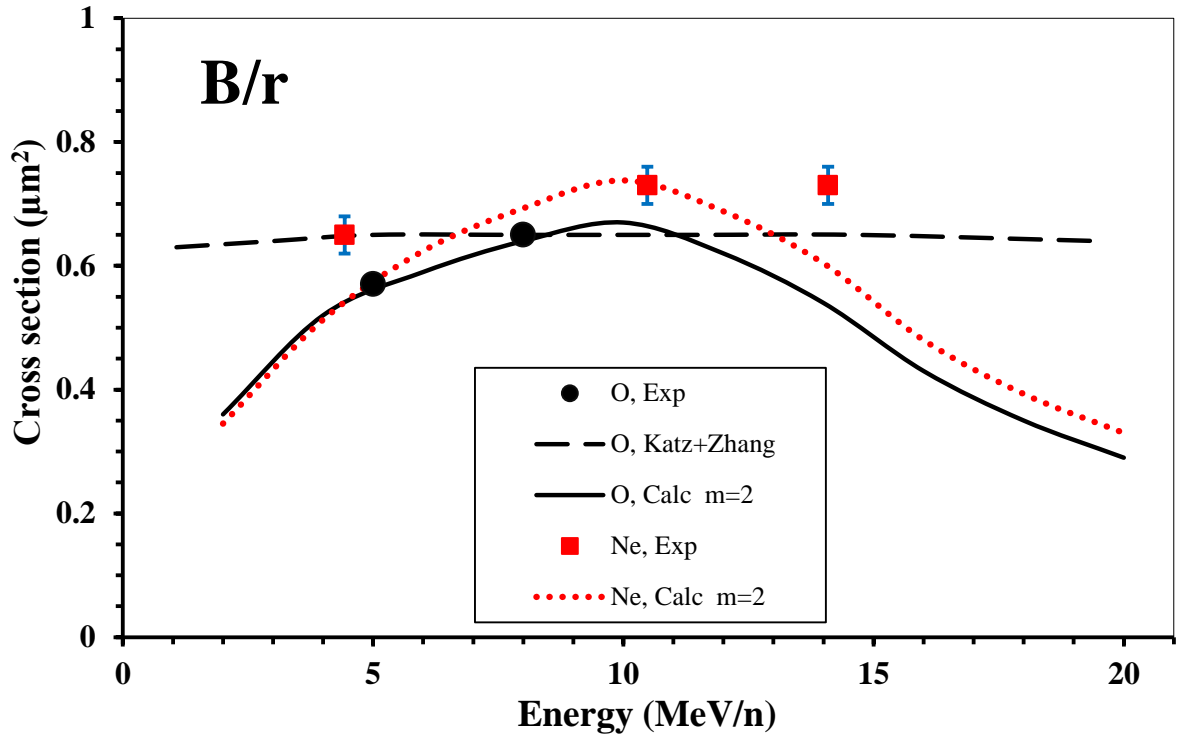

Fig.10

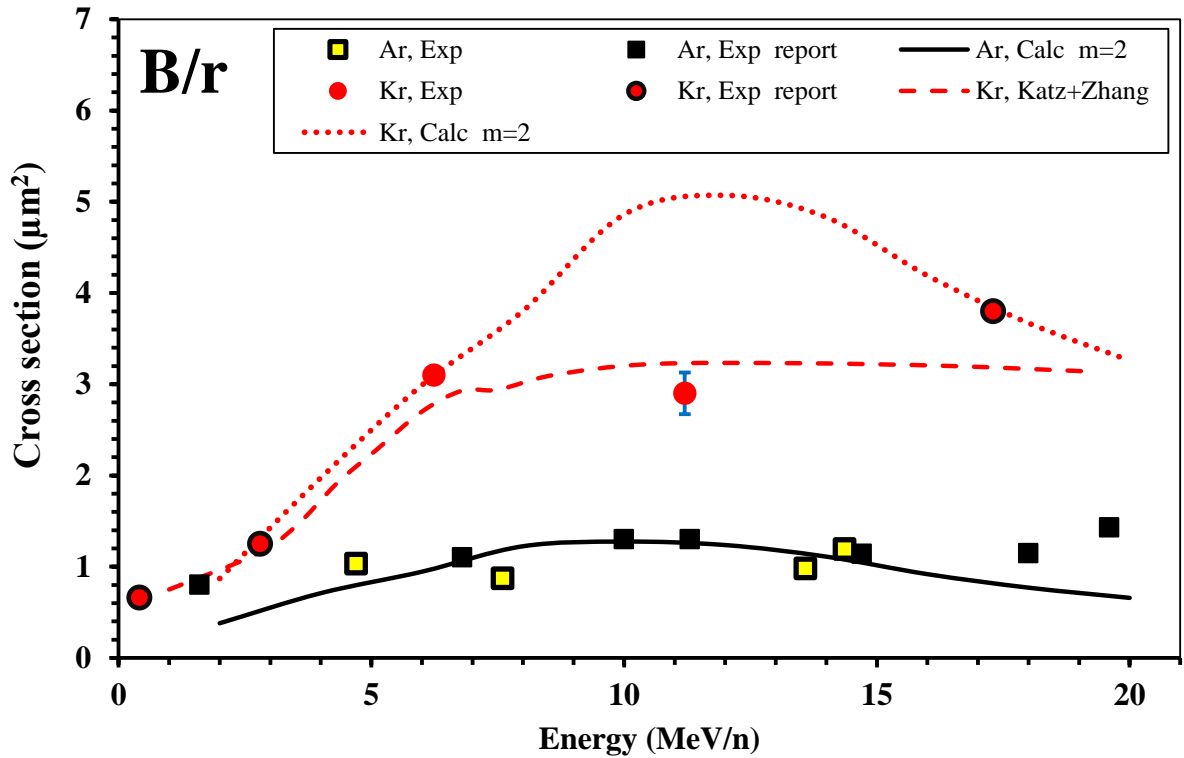

Fig.11

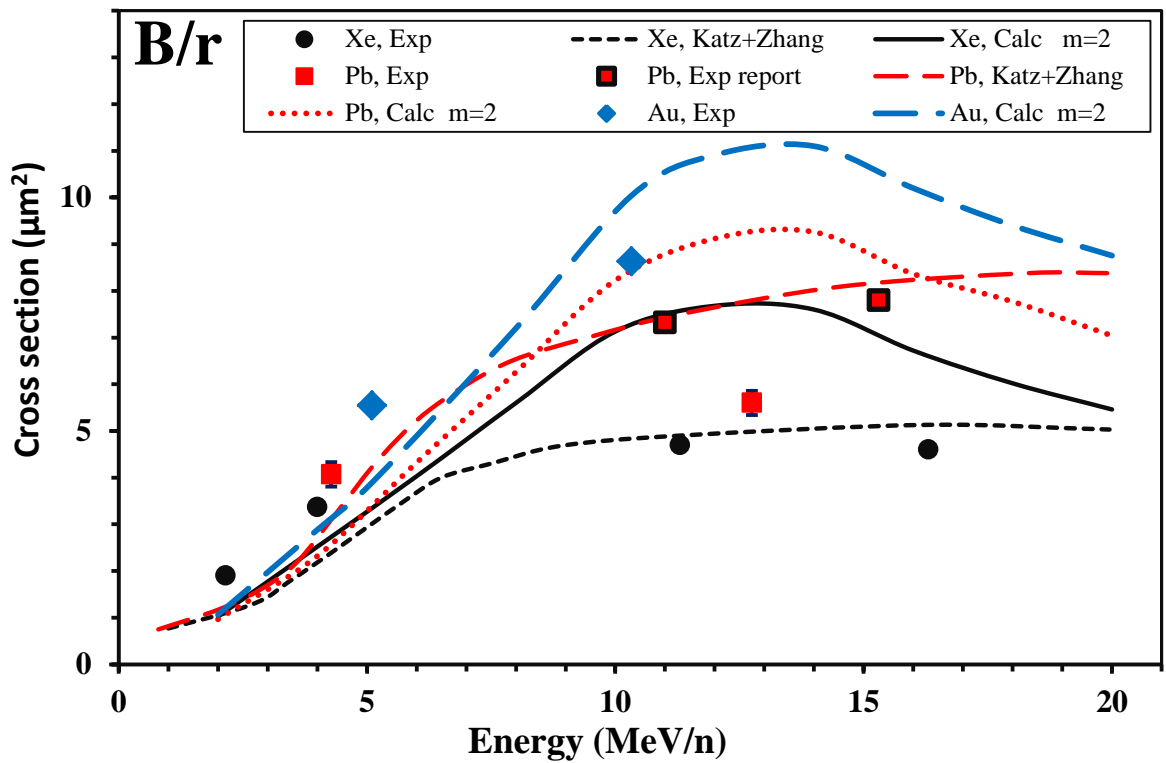

Fig. 12

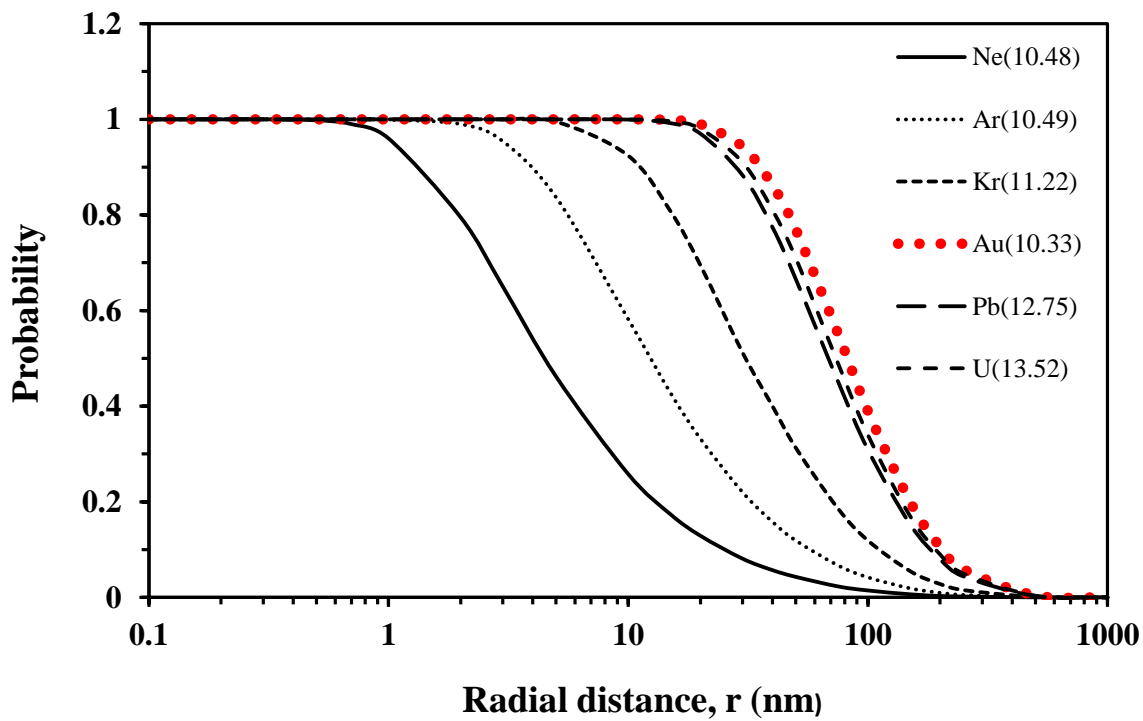